\documentstyle[12pt,aasms4]{article}

\begin{document}

\title{GONG p-mode frequency changes with solar activity}

\author{A. Bhatnagar,\altaffilmark{1} Kiran Jain,\altaffilmark{2} and S. C. Tripathy\altaffilmark{3}}
\affil{Udaipur Solar Observatory, A unit of  Physical Research Laboratory\\
P.O. Box No.198, Udaipur - 313 001, INDIA \\}
\authoremail{arvind@uso.ernet.in}
\lefthead{Bhatnagar, Jain and Tripathy}
\righthead{Solar activity and GONG frequencies}
\altaffiltext{1}{arvind@uso.ernet.in}
\altaffiltext{2}{kiran@uso.ernet.in}
\altaffiltext{3}{sushant@uso.ernet.in}
\begin{abstract}

We present a correlation analysis  of GONG 
p-mode frequencies with nine solar activity indices for
the period 1995 August  to 1997 August. This study includes
spherical harmonic degrees in the range 2 to 150 and the
frequency range of 1500-3500~$\mu$Hz. Using three statistical
tests, the measured mean frequency shifts show strong to good
correlation with activity indices. A decrease of 0.06~$\mu$Hz in
frequency, during the descending  phase of solar cycle 22
and an increase of 0.04 $\mu$Hz in the ascending phase of solar
cycle 23 is observed.  These results  provide the first evidence 
for change in p-mode frequencies
around the declining phase of cycle 22 and beginning of new
cycle 23. This analysis further confirms that the temporal behavior 
of the solar frequency shifts closely follow the phase of the solar activity cycle.

\end{abstract}

\section{INTRODUCTION}

Solar surface phenomena such as sunspots, magnetic field, 11-year solar cycle, and associated 
activities are the consequence of dynamical processes occurring inside the Sun. The fundamental 
parameters responsible for most of the activities are the magnetic and velocity fields, thus the 
acoustic mode (p-mode) frequencies could be influenced by the solar activity. Hence, a relation 
between the solar frequencies and activity indices may throw some light on the changes occurring 
deep in the solar interior and provide clues to the mechanism of solar cycle. The first evidence that 
the frequencies of the p-modes change with the solar cycle, came from the observations of low 
degree ($\ell$) modes. From the low-$\ell$ ACRIM data a decrease in the frequencies of 
0.42~$\mu$Hz between 1980 and 1984 was reported by Woodard and Noyes (1985). It was 
suggested by  \cite{kun88} that the observed variation in frequencies is due to the changes in the 
internal temperature structure and are correlated with the variation in the surface 
temperature. Subsequently, \cite{gol91} theoretically showed that the variations in the 
solar p-mode eigen frequencies are related to the perturbations in the magnetic flux at the sun's 
surface. Observations made during the rising phase of the solar cycle 22 by the Big Bear Solar 
Observatory (BBSO) group found an increase  of about 0.4~$\mu$Hz in mean frequency during the period 1986-89 
(\cite{lw90}).  Further, it was shown that the frequencies varied on a time 
scale as short as 3 weeks with the absolute value of the magnetic field strength over the visible 
solar disk (Woodard et al. 1991). These results confirmed the theoretical work by 
\cite{gol91}. Now, there is growing evidence that the mode frequencies vary with solar activity. 
\cite{bb93} using the High Altitude Observatory's (HAO) Fourier Tachometer data for $\ell$ 
between 20 and 60 in the frequency range of 2600-3200 ~$\mu$Hz showed that the p-mode 
frequency shifts correlate remarkably well with six solar activity indices during 1984 October 
through 1990 November. For the low degree modes ($\ell~<~$3) and during different time epochs, 
several groups (\cite{ang92}, R\'{e}gulo et al. 1994, \cite{jim98}) have 
shown that the frequency shifts are well correlated with solar activity cycle. 

It has also been shown that the frequency splitting coefficients vary with solar cycle 
(Kuhn 1988,~\cite{wod93} ). It was pointed out that the even-order coefficients are sensitive to 
the solar cycle or to the latitude-dependent properties while the odd-order coefficients reflects the 
advective, latitudinally symmetric part of the perturbations caused by rotation. Recently, using the 
GONG data, \cite{and98} have found evidence for small shifts in the central frequencies and splitting 
coefficients.  From the solar oscillation data obtained from Michelson Doppler Imager on 
board SOHO, during 1996 May 1 through 1997 April 25, \cite{dzi98} also detected a significant 
trend in the  splitting coefficients. 

Motivation for the present work arose due to the availability of high precision frequencies from the 
GONG network, and for extending  the earlier studies to include the declining phase of cycle 22 
and the beginning of cycle 23. In this study, we also look for a possible correlation between the p-
mode frequency shifts and nine solar activity indices representing the photospheric, chromospheric 
and coronal activities.

\section{OSCILLATION  DATA  AND  SOLAR  ACTIVITY  INDICES}

The GONG data (\cite{hil96}) used in this study  consist of 8 data sets covering a period of two 
years from 1995 August 23 to 1997 August 11, in the frequency range between 1500 to
3500~$\mu$Hz and $\ell$ from 2 to 150. This period covers the GONG months (GM) 4 to 9 (1995 
August 23 to 1996 March 25) each of 36 days, GM 12-14 (1996 June 6-September 21) and GM 
21-23 (1997 April 26-August 11) of 108 days each. The period from 1995 August through 1996 
September covers the declining phase of the solar cycle 22, whereas the period from 1997 
April-August refers to the beginning of cycle 23.

For calculating the frequency shifts, we have used the frequency of GM 12-14 as a reference. The 
choice of this standard frequency has been made in view of two reasons; (a) that this datum lies in 
the middle of the period covered in this study, and (b) that it represents the period of the minimum 
solar activity. The mean frequency shift is found from the relation:
\begin{equation}
\delta\nu(t)  =  {\sum_{n\ell}\frac{Q_{n\ell}}{\sigma_{n\ell}^2}\delta\nu_{n\ell}(t)}/{\sum_{n\ell}\frac{Q_{n\ell}}{\sigma_{n\ell}^2}}
\end{equation}

where $Q_{n\ell}$ is the inertia ratio as defined by \cite{jcd91},  $\sigma_{n\ell}$ is the
error in frequency measurement, as provided by  GONG  and $\delta\nu_{n\ell}$(t) is the change in 
the measured frequencies for a given $\ell$, and radial order, $n$. We have neglected those 
mode frequencies for which $\sigma_{n\ell}~>$~0.1 $\mu$Hz, to avoid large errors in calculating 
the mean shifts. The resulting mean weighted frequency shift $\delta\nu$ is plotted against the 
GONG months for four different $\ell$ ranges (2~$\le~\ell~\le$~19, 20~$\le~\ell~\le$~60, 
61~$\le~\ell~\le$~150, and 2~$\le~\ell~\le$~150) and is shown in Figure~1. For the interval GM 4 
to 9, the frequency shift shows a systematic decrease of about 0.06~$\mu$Hz for all $\ell$ values 
while for GM 9 to 23, the mean shift shows an increase of 0.04~$\mu$Hz, indicating that the solar 
oscillation frequency varies with solar activity. 
\placefigure{fig1}
We have correlated the mean frequency shifts with different solar activity indices: the International 
sunspot number, $R_I,$  obtained from the Solar Geophysical Data (SGD); KPMI, Kitt Peak 
Magnetic Index from Kitt peak full disk magnetograms (\cite{har84}); SMMF, Stanford Mean 
Magnetic Field from SGD;  MPSI, Magnetic Plage Strength Index from Mount Wilson 
magnetograms (\cite{ulr91});  FI, total flare index from SGD and \cite{ata99}; He~I, equivalent width of Helium~I 
10830$\AA$ line, averaged over the whole disk from Kitt peak;  Mg~II, core-to-wing ratio of 
Mg~II line at 2800$\AA$ from SUVSIM (SGD); F$_{10}$, integrated radio flux at 10.7 cm from 
SGD;  Fe~XIV, Coronal line intensity at 5303$\AA$ from SGD. A mean value was computed for 
each activity index over the interval corresponding to the same GONG month.

\section{ANALYSIS  AND  RESULTS}

To study the relative variation in frequency shift $\delta\nu$ with activity index $i$, we assume a 
linear relationship of the form:
\begin{equation}
\delta\nu  =  ai + b ,
\end{equation}
where $\delta\nu$ includes the mean error in the measured frequencies. The slope, $a$ and the 
intercept, $b$ are obtained by performing a linear least square fit. Figure~2 shows a
typical plot of our analysis for  all modes between 2 to 150. The solid line represents the 'best fit' 
regression line and shows that the data is consistent with the assumption of a linear relationship. 
The bars represent 1$\sigma$ error in fitting. We also carried out $\chi^2$-test which takes into account the 
statistical uncertainties of each of the frequency shift. As there is no reliable 
estimate available of the uncertainties in the measurement of activity indices, those are not included 
in the fitting.
\placefigure{fig2}
\placetable{tb1}

	The linear relationship given in equation~(2) is further tested by calculating the parametric 
Pearson's coefficient, $r_p$, and the non-parametric Spearman's rank correlation coefficient, $r_s$
along with their two-sided significance, $P_p$, and $P_s$ respectively, for all the $\ell$ ranges. As 
a typical example, these parameters are given in Table~1, for the $\ell$ range of 2 to 150. From the 
table, it is clear that a positive correlation exists for all the activity indices. We note that the best 
correlation is obtained for Mg~II index, while F$_{10}$ and Fe~XIV indices show good 
correlation. SMMF show poor correlation, perhaps due to large gaps in the available data. 
Normally, we expect better correlation should exist between the surface magnetic activity and 
$\delta\nu$, as this is the fundamental parameter for all solar activity.  However, from Table~1, we 
notice that the radiative indices ($F_{10}$, Mg~II and Fe~XIV) show better correlation as 
compared to magnetic field indices. Backmann and Brown (1993) using the HAO data had obtained a 
similar result. The reason for such a difference in the level of correlation needs further 
investigation.

	In order to investigate the degree dependence of frequency shifts with activity indices, we 
carried out a statistical analysis for three different ranges of $\ell$. It is seen that the fitting 
parameters, $a$ and $b$ do not show significant variation with $\ell$. The lowest $\chi^2$
values are found for 2~$\le~\ell~\le$~19, while for 20~$\le~\ell~ \le$~60 and 
61~$\le~\ell~\le$~150 the $\chi^2$ values are  higher. It is further noted that the correlation 
coefficients, $r_p$, and $r_s$ for all $\ell$ ranges have values between 0.75 to 1.0 for all solar 
indices except for He~I and SMMF. Comparing $r_p$ and $r_s$ values for all activity indices, it is
noted that for 2~$\le~\ell~\le$~19 these values are systematically less, indicating a week 
correlation as compared to the other $\ell$-ranges. Thus, there appears to be some evidence that the 
frequency shift depends on $\ell$. 
\placetable{tb2}

	Comparing our results for 20~$\le~\ell~\le$~60 with those of \cite{wod91}, and 
\cite{bb93} for modes of similar degrees (Table~2), we find that the magnitude of slope $a$, is 
comparable in all the three cases for the KPMI index, the only index common in all the three 
analysis. However, a small difference of 0.003~$\mu$Hz between the GONG and BBSO data and 
0.001~$\mu$Hz between the GONG and HAO data is noticed. This difference may be due to 
different frequency intervals and phase of the solar cycle. A detailed comparison between BBSO 
and HAO results was made by \cite{reg94}.  The other four indices considered in HAO and our 
work show similar correlation.
\placefigure{fig3}
	
	The temporal behavior of the solar frequency shifts and the mean monthly sunspot number 
$R_z$ is shown in Figure~3. The diamonds represent the HAO data for the period 1984 October to
1990 October, while the triangles indicate the GONG data for the period 1995 August to 1997 
August. A polynomial of order $n$ of the form
\begin{equation}
\delta\nu  =  a_0 + \sum_{i=1}^n a_id^i
\end{equation}
is fitted separately to each set of frequency shift $\delta\nu$ since the reference frequencies for 
HAO and GONG data are different.  Here, $a_0$, and $a_i$ are the fitting coefficients
and $d$ the number of days after 1981 January 1. The best fit for HAO data is obtained with a 
polynomial of order 4, while a polynomial of order 2 fits the GONG data. The fitting coefficients  
for HAO data are; $a_0$~=~0.557, $a_1$~=~1.583x10$^{-4}$, $a_2$~=~$-$1.107x10$^{-6}$,   
$a_3$~=~5.839x10$^{-10}$, $a_4$~=~$-$8.282x10$^{-14}$, and for GONG data are; 
$a_0$~=~10.948, $a_1$~=~$-$3.847x10$^{-3}$, $a_2$~=~3.369x10$^{-7}$.  From this plot it is evident that 
the mean frequency shifts systematically follow the solar activity cycle 22 and the beginning of 
cycle 23.

	We conclude that a positive and linear correlation exists between the frequency shift and the 
nine solar activity indices. It is observed that the radiative indices show better correlation as 
compared to the magnetic indices. This analysis shows a 
decrease of 0.06~$\mu$Hz in weighted frequency during the declining phase of solar cycle 22 and 
an increase of 0.04 $\mu$Hz during the ascending phase of cycle 23. Our finding confirms that the 
temporal behavior of the solar frequency shifts, closely follow the phase of the solar activity cycle 
and we conjecture that the same trend will continue during cycle 23.

\acknowledgements 
We thank H.M. Antia for his critical remarks and helpful discussions. We also thank T. Ata\c{c}, L. 
Floyd, and R.K. Ulrich for supplying us the Flare index, Mg II (v19r3), and MPSI data 
respectively. The authors are thankful to the referee for critical suggestions. This work utilizes data 
obtained by the Global Oscillation Network Group project, managed by the National Solar 
Observatory, a Division of the National Optical Astronomy Observatories, which is operated by 
AURA, Inc. under cooperative agreement with the National Science Foundation. The data were 
acquired by instruments operated by Big Bear Solar Observatory, High Altitude Observatory, 
Learmonth Solar Obsrvatory, Udaipur Solar Observatory, Instituto de Astrophsico de Canaris, and 
Cerro Tololo Interamerican Observatory. NSO/Kitt Peak magnetic, and Helium measurements used 
here are produced cooperatively by NSF/NOAO; NASA/GSFC and NOAA/SEL.  This work is 
partially supported under the CSIR Emeritus Scientist Scheme and Indo-US collaborative 
programme--NSF Grant INT-9710279.

\newpage
\figcaption[ab1.ps]{Mean frequency shifts for p-modes weighted by the mode inertia in the 
frequency range of 1500-3500 $\mu$Hz for different ranges of $\ell$; 
2~$\le~\ell~\le$~19 (dotted line), 20~$\le~\ell~\le$~60 (dashed), 
61~$\le~\ell~\le$~150 (dash-dot), and 2~$\le~\ell~\le$~150 (solid).  Error bars
represent mean error in shift.\label{fig1}}

\figcaption[ab2.ps]{Comparison of weighted mean frequency shift with nine different 
activity indices in the $\ell$ range of 2 to 150. These are (a) sunspot number,
 (b) Kitt Peak magnetic index (Gauss), (c) 10.7 cm flux (sfu), (d) He~I 10830
  $\AA$ equivalent width (m$\AA$), (e) Stanford mean magnetic field ($\mu$Tesla),
  (f) Magnetic plage strength index (Gauss), (g) Total flare index, (h) Coronal line
  intensity (10$^{16}$W/sr), (i) Mg~II core-to-wing ratio.  The solid line represents the
linear fit. The error bars indicate 1$\sigma$ error of the fitting.\label{fig2}} 

\figcaption[ab3.ps]{ The top panel shows the time variation of frequency
shifts for the HAO~(diamond) and GONG (triangles) data for a
total period from 1984 to 1997. 
The best fit for GONG data is obtained with a polynomial of order 2
while the HAO data is best represented by a polynomial of order
4. For comparison, the bottom panel shows the time variation of the 
mean monthly sunspot number $R_z$. Error bars for GONG data are not clearly
visible as these are very small.\label{fig3}}

\newpage
\begin{deluxetable}{llccccccc} 
\tablecaption{Fitting and correlation statistics for 2 $\le \ell \le$ 150 \label{tb1}}
\tablewidth{0pt}
\tablehead{
\colhead{Activity index}&\colhead{slope $a$} & \colhead{intercept $b$ ($\mu$Hz)} 
&\colhead{$\chi^2$} &\colhead{$r_p$}&\colhead{$P_p$}&\colhead{$r_s$}&\colhead{$P_s$}}
\startdata
R$_I$&0.002 $\pm$ 0.0001 $\mu$Hz&$-$0.034 $\pm$ 0.001&344&0.82&0.023&0.93&0.002\\
KPMI&0.029 $\pm$ 0.001 $\mu$Hz/G&$-$0.221 $\pm$ 0.008&461&0.81&0.025&0.71&0.071\\
F$_{10}$&0.005 $\pm$ 0.0002 $\mu$Hz/sfu&$-$0.345 $\pm$ 0.011&144&0.93&0.002&0.93&0.002\\
He I  &0.010 $\pm$ 0.0004 $\mu$Hz/m\AA&$-$0.449 $\pm$ 0.017&480&0.76&0.049&0.61&0.148\\
SMMF&0.0004 $\pm$ 0.0002 $\mu$Hz/$\mu$T&$-$0.009 $\pm$ 0.002&1130&0.16&0.729&0.21&0.645\\
MPSI&0.191 $\pm$ 0.006 $\mu$Hz/G&$-$0.031 $\pm$ 0.0009&266&0.87&0.011&0.93&0.002\\
FI&0.023 $\pm$ 0.0008 $\mu$Hz&$-$0.017 $\pm$ 0.0005&373&0.83&0.020&0.93&0.002\\
Fe XIV&0.012 $\pm$ 0.0004\tablenotemark{*} &$-$0.048 $\pm$ 0.001&128&0.93&0.006&0.94&0.005\\
Mg II&38.210 $\pm$ 1.172 $\mu$Hz&$-$9.712 $\pm$ 0.297&68&0.97&0.003&0.90&0.005\\
\enddata
\tablenotetext{*}{$\mu$Hz sr/10$^{16}$W}

\end{deluxetable} 
\clearpage
\newpage
\begin{deluxetable}{lccccccc} 
\tablecolumns{8}
\tablewidth{0pt}
\tablecaption{Comparison between GONG and HAO results \tablenotemark{a}\label{tb2}}
\tablehead{
\colhead{}& \multicolumn{3}{c}{GONG analysis}&\colhead{}&\multicolumn{3}{c}{HAO analysis}\\
\cline{2-4} \cline{6-8}\\
\colhead{Activity Index}&\colhead{$a$} & \colhead{$b$} 
&\colhead{$\chi^2$}&\colhead{} &\colhead{$a$}&\colhead{$b$}&\colhead{$\chi^2$}}
\startdata
KPMI\tablenotemark{*}&0.024 &$-$0.182&216&&0.023&$-$0.306&434\\

F$_{10}$&0.004 &$-$0.307&60&&0.003&$-$0.345&153\\

He I&0.009&$-$0.410&170&&0.011&$-$0.667&184\\

MPSI&0.171&$-$0.031&94&&0.136&$-$0.168&166\\

Mg II&33.986&$-$8.643&22&&20.96&$-$5.700&134\\
\enddata
\tablenotetext{a}{Units of $a$ and $b$ are same as given in Table~1}
\tablenotetext{*}{for BBSO, $a$~=~0.027 $\mu$Hz/G}
\end{deluxetable}

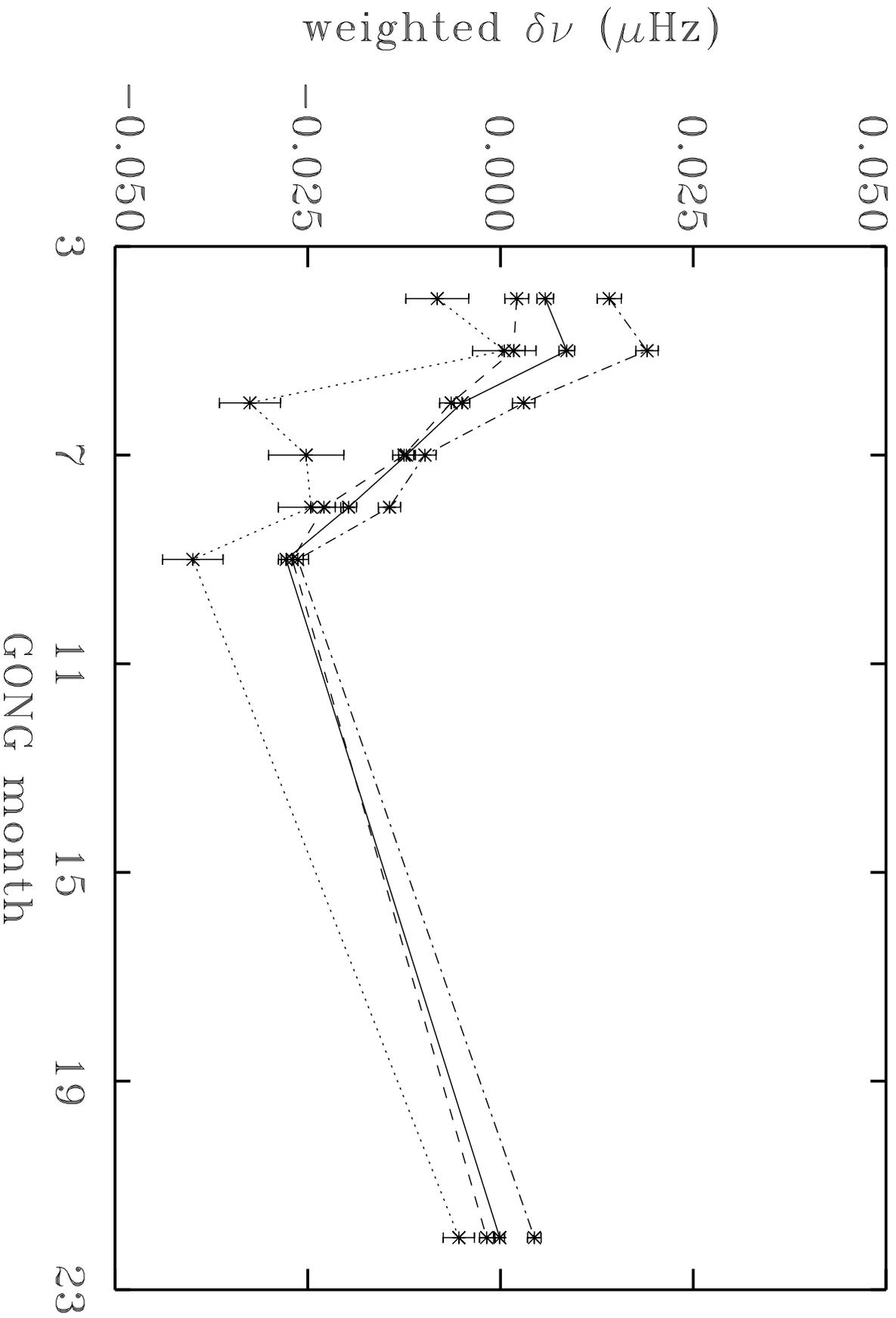
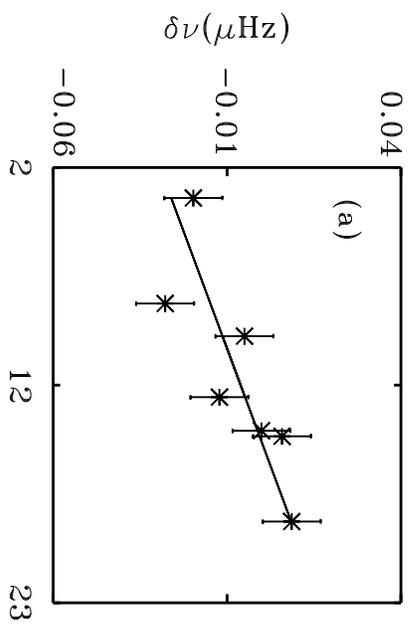
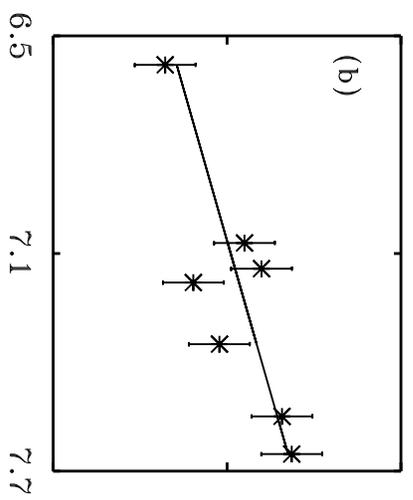
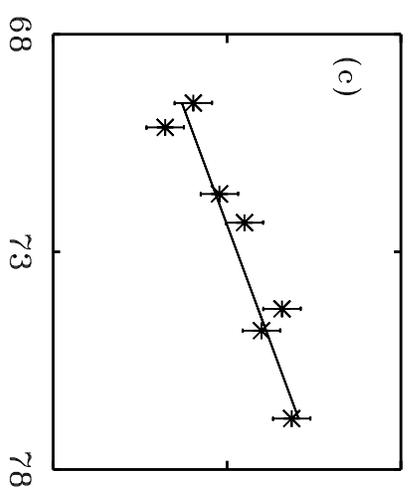
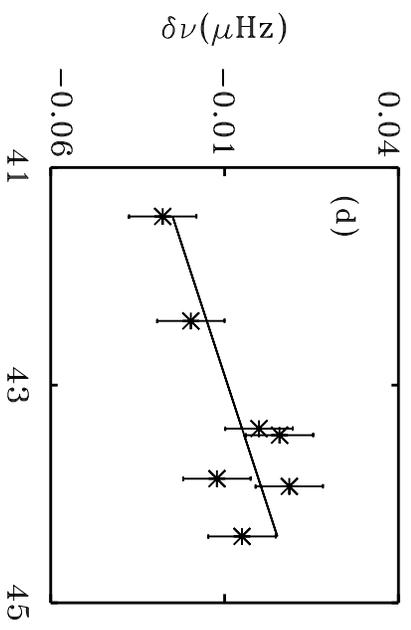
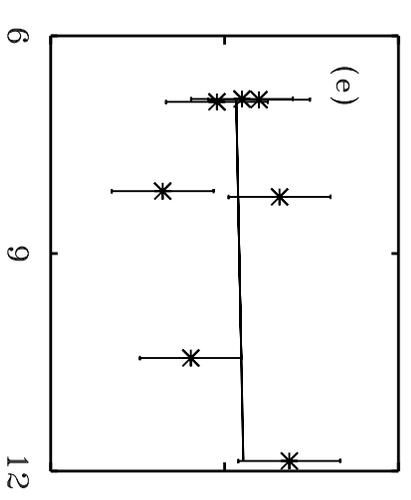
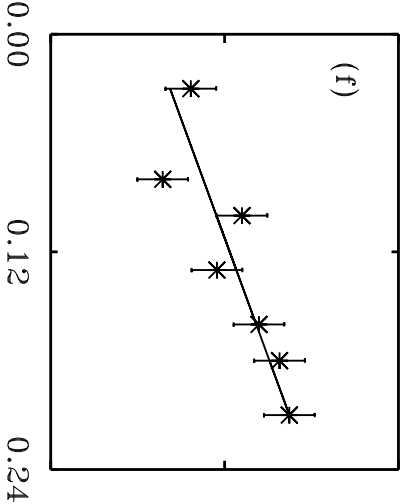
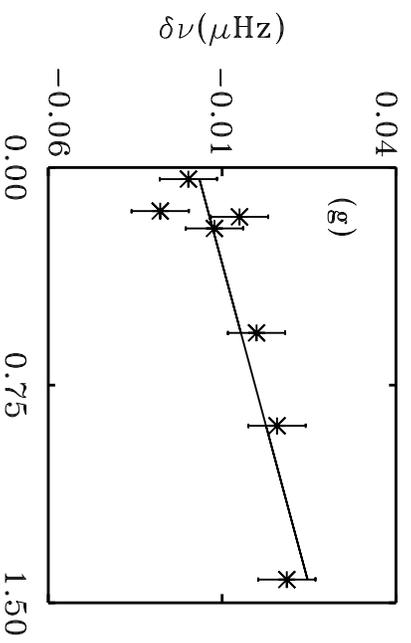
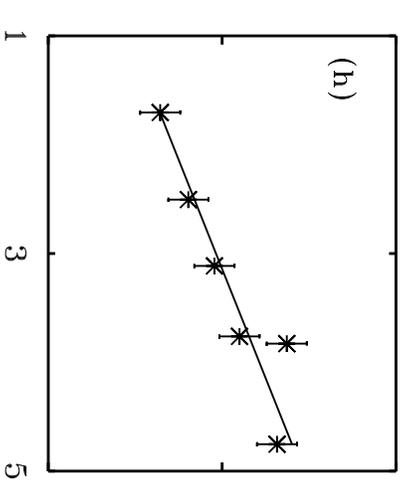
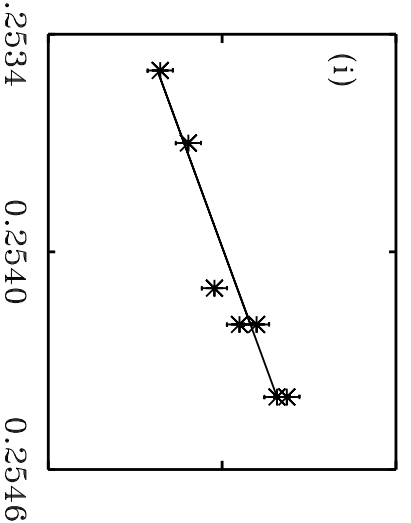
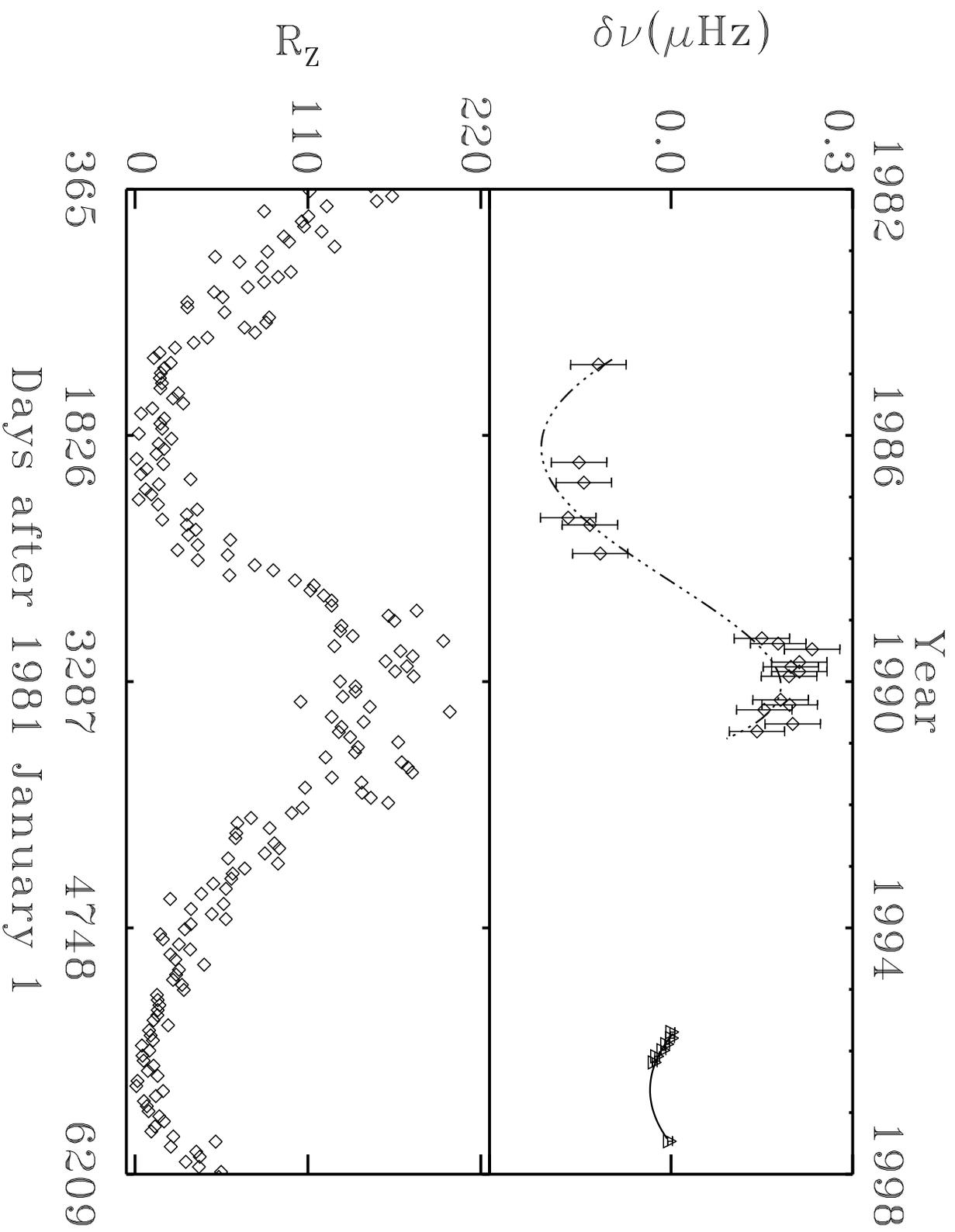
\end{document}